\documentclass[12pt]{article}
\usepackage{graphicx,aaspp4}
\def\Msun {\hbox{$M_\odot$}}

\def\J {$J$}

\def\kms {\hbox{${\rm km\, s}^{-1}$}}

\def\arcsec {\hbox{$^{\prime\prime}$}}

\def\HTW {\hbox{\rm H$_2$}}    

\def\farcs{\hbox{$.\!\!^{\prime\prime}$}}

\newbox\grsign      \setbox\grsign=\hbox{$>$}
\newdimen\grdimen   \grdimen=\ht\grsign                                       
\newbox\simlessbox  \setbox\simlessbox =\hbox{\raise.5ex\hbox{$<$}\llap
  {\lower.5ex\hbox{$\sim$}}}\ht2=\grdimen\dp2=0pt 
\def\simless {\mathrel{\copy\simlessbox }}            
\def\vinf   {$v_{\infty}$}    
\slugcomment{Astrophysical Journal (Letters), accepted}
 
\begin{document}
 
\title{A SUBMILLIMETER HCN LASER IN IRC+10216}
 
\author{Peter Schilke\altaffilmark{1}, David M. Mehringer\altaffilmark{2,3},
\&\ Karl M. Menten\altaffilmark{1}}

\altaffiltext{1}{Max-Planck-Institut f{\"u}r Radioastronomie, 
Auf dem H{\"u}gel 69, Bonn, D-53121, Germany}

\altaffiltext{2}{California Institute of Technology, MS 320-47, Pasadena,
CA 91125}

\altaffiltext{3}{Present address: University of Illinois, Department of 
Astronomy, 1002 W. Green St., Urbana, IL 61801}

\begin{abstract}
We report the detection of a strong submillimeter wavelength HCN laser
line at a frequency near 805 GHz toward the carbon star IRC+10216.  
This line, the \J=9--8 rotational transition 
within the $(04^00)$ vibrationally
excited state, is one of a series of HCN laser lines that were first detected
in the laboratory in the early days of laser spectroscopy. Since its 
lower energy level is 4200 K above the ground state, the laser emission must
arise
from the inner part of IRC+10216's circumstellar envelope. To better
characterize this environment, we observed other, thermally emitting,
vibrationally excited HCN lines and find that they, like the laser line, 
arise
in a region of temperature $\approx 1000$ K that is located
within the dust formation radius; this conclusion is 
supported by the linewidth 
of the laser. The $(04^00)$, \J=9--8 laser might be chemically pumped
and may be the only known laser (or maser) that is excited both in the
laboratory and in space by a similar mechanism. 
\end{abstract}
 

\section {Introduction}

Rotational lines from vibrationally excited states of various
molecules are useful probes of the hottest and densest regions of
circumstellar envelopes around asymptotic giant branch (AGB) stars.
For example, intense maser emission from vibrationally excited SiO (up
to the $v=4$ state; Cernicharo, Bujarrabal, \&\ Santar{\'e}n 1993) and
H$_2$O (in the $\nu_2$ bending mode; Menten \&\ Young 1995) is a
ubiquitous phenomenon in oxygen-rich red giants and supergiants.
Interferometric observations show that these masers arise from a
region of thickness a few stellar radii that is located between the
photosphere and the inner edge of the dust formation zone (Diamond et
al. 1994; Greenhill et al. 1995; see Habing 1996 for a recent review).

In carbon-rich AGB stars, (sub)millimeter-wavelength vibrationally
excited rotational lines have been observed from, among others, the
HCN, CS, and SiS molecules in particular toward the extreme carbon
star IRC+10216 (CW Leo) (Turner 1987a,b; Ziurys \&\ Turner 1986). This
object is due to its proximity ($D \approx 140$ pc) and high mass-loss
rate ($\dot M \approx 2\times10^{-5}$ \Msun~yr$^{-1}$, Crosas \&\ 
Menten 1997; Groenewegen, van der Veen, \&\ Matthews 1998) one of the
most prolific and best-studied molecular line sources in the sky (see,
e.g., Avery et al. 1992; Groesbeck, Phillips, \&\ Blake 1994;
Kawaguchi et al. 1995; Cernicharo, Gu{\'e}lin, \&\ Kahane 1999).
Recent millimeter-wavelength interferometric observations have shown
that the vibrationally excited emission from all three of the
molecules mentioned above arises from the innermost parts of
IRC+10216's circumstellar envelope, i.e. from a region with a radius
of $\approx 5$--10 stellar radii, $r_\star$ (Lucas \&\ Guilloteau
1992; Lucas \&\ Gu{\'e}lin 1999); we assume a value of $0\farcs 023$
for $r_\star$ (Danchi et al. 1994). Therefore, vibrationally excited
lines are interesting probes of the hot ($T \sim 1000$ K), dense
regions in which dust grains form (see, e.g., Winters, Dominik, \&\ 
Sedlmayr 1994; Danchi et al. 1994; Groenewegen 1997).

In the past, only two molecular lines, both from HCN, have been found
to show strong maser emission in carbon stars. One of these, the 177
GHz \J=2--1 transition from the $(01^{1_c}0)$ excited bending mode was
detected toward IRC+10216 by Lucas \& Cernicharo (1989). The high flux
density ($\approx 400$ Jy), asymmetric line profile, and time
variability (Cernicharo et al. 1999) clearly prove that this line is
masing. In this Letter, we report the detection of submillimeter
laser\footnote{In the laboratory spectroscopy literature, the term
  laser is used for stimulated emission in the far-infrared and
  submillimeter regimes, i.e. between microwave and optical
  wavelengths, in particular for the HCN line discussed here.  We have
  followed this convention (see also Strelnitski et al. 1996), instead
  of using the term maser, which is common in radio astronomy.}
emission in the \J=9--8 rotational line within the highly excited
$(04^00)$ vibrational state of HCN at a frequency near 805 GHz.

\section{Observations}
The observations were made in 1998 February as part of a systematic
study of IRC+10216's submillimeter spectrum with the 810~GHz receiver
(Kooi et al. 1998) of the Caltech Submillimeter Observatory (CSO) 10.4
m telescope on Mauna Kea, Hawaii.  Typical single sideband system
temperatures were about 5000~K under excellent weather conditions,
i.e. time periods when the precipitable water vapor content was below
1 mm.
Pointing was checked by observing the CO(7--6) line toward IRC+10216
itself, resulting in a pointing accuracy of $\approx 2$\arcsec.  At
805 GHz the beam width is $9''$ (FWHM) and the main beam efficiency
was found to be 0.3 from observations of Mars.  The spectrometer and
observing procedure are described in Menten \&\ Young (1995). We
estimate our absolute calibration uncertainties to be $\approx 30$\%.
Additional observations of HCN \J=3--2, 4--3, and 8--7 rotational
lines within different vibrational states at frequencies near 266,
355, and 709~GHz were made in 1998 April with the same telescope.

\begin{figure}[htbp]
  \begin{center}
\includegraphics[bb=170 470 460 700]{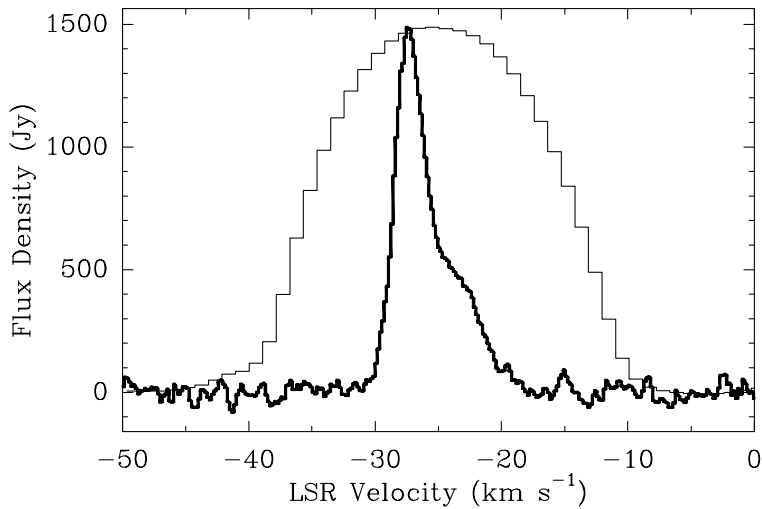}
\caption{Spectrum of the $(04^00)$, \J=9--8 laser line
  is shown as the bold line. The very narrow linewidth and asymmetric
  shape are conspicuous when compared to a spectrum of the HCN \J=4--3
  transition from the vibrational ground state, which is overlaid as
  the thin line. The flux density scale is appropriate for the
  $(04^00)$, \J=9--8 line only.}
    \label{fig:fig1}
  \end{center}
\end{figure}

\section{Results and Discussion}
\subsection{A HCN submillimeter laser}
At millimeter wavelengths IRC+10216 has a rich molecular spectrum with
emission from more than 50 species arising from a compact central
region and/or from hollow shells centered on the star. Radii and
thicknesses of these shells depend on the excitation requirements of
the molecule in question and on the chemistry.  Generally speaking,
since submillimeter transitions need much higher densities and
temperatures for their excitation than lower frequency lines, the
submillimeter spectrum of IRC+10216 is considerably sparser than its
millimeter spectrum. In particular, lines from long chain molecules
and other radicals, which reside in shells of modest temperature and
density (see, e.g., Lucas \&\ Gu{\'e}lin 1999), are largely absent and
most of the detected lines can be assigned to simple diatomic or
triatomic species emitted from the inner envelope.  Moreover, almost
all of the submillimeter lines observed toward IRC+10216 are readily
assigned to known carriers (Groesbeck et al.  1994), whereas in the
mm-range many (weak) lines remain unidentified (Cernicharo et al.
1999).

We were thus surprised to discover a strong spectral line at a
frequency of $\approx 804.751$~GHz (Fig.~1), which we could not
readily identify in available line catalogs (e.g., Pickett et. al.
1998). The line's frequency was verified by checking its sideband
assignment by shifting the local oscillator frequency.  With a total
width $\simless 10$~\kms (FWZP) the emission covers a much smaller
velocity interval than most lines observed toward IRC+10216, which
have full widths of 29 \kms, i.e. twice the terminal outflow velocity,
\vinf. This indicates that the emission arises from a region in which
the outflow has not yet reached its terminal velocity and/or that the
line shows laser action. The latter is also suggested by the
asymmetric line profile which is distinctly different from the
symmetric parabolic, rectangular, or double ``horn''-shaped profiles
expected and commonly observed in the IRC+10216 envelope.

A literature search resulted in the identification of the line with
the \J=9--8 transition of HCN within the $\nu_2=4$ vibrationally
excited bending mode\footnote{The linear triatomic HCN molecule has
  three vibrational states, one bending mode $\nu_2$, with $\nu_2=1$
  being $\approx 1000$~K above ground, and two stretching vibrations:
  $\nu_1$, corresponding to the CN stretch, where $\nu_1=1$ is 3000~K
  above the ground state, and $\nu_3$, the CH stretch, with $\nu_3=1$
  being 4700~K above ground.  [We follow the notation of Herzberg
  (1945) and note that various papers differ in their usage of the
  $\nu_1$, $\nu_2$, and $\nu_3$ quantum numbers.]  The bending mode
  $\nu_2$ is doubly degenerate; for $\nu_2\ne 0$ this degeneracy is
  lifted and the levels are split by rotation-vibration interaction.
  A new quantum number, $\ell$ (where $\ell=\nu, \nu-2, \ldots -\nu$)
  is needed to describe the system.  Overtones and combination bands
  also exist.  From here on, the notation ($\nu_1 \nu_2^\ell \nu_3$)
  is used.} (0$4^0$0), whose lower (\J=8) level is 4163 K above the
ground state and whose frequency was measured by Hocker \&\ Javan
(1967) as 804.7509 GHz with an estimated uncertainty of 1 MHz.  Most
interestingly, this line was among the first submillimeter laser
transitions detected in the laboratory (Mathias, Crocker, \&\ Wills
1965) and is part of the extensively studied (1$1^1$0)/(0$4^0$0)
coriolis-perturbed system around the \J = 9 to 11 rotational levels in
these states (Gebbie, Stone, \&\ Findlay 1964; Lide \&\ Maki 1967).
These lasers, which are produced in gas discharges, were intensely
studied in the early years of molecular laser research (for reviews
see, e.g., Chantry 1971; Kneub{\"u}hl \&\ Sturzenegger 1980;
Pichamuthu 1983).

\begin{figure}[htbp]
  \begin{center}
    \includegraphics[scale=0.5,angle=-90]{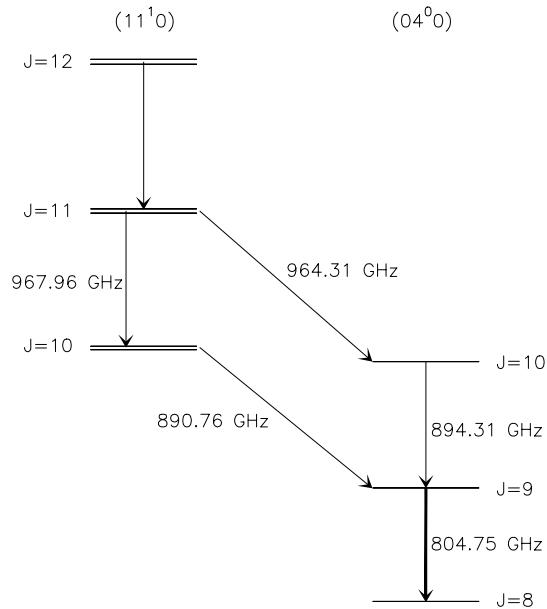}
\caption{Excerpt from the level diagrams of 
  the $(11^10)$ and $(04^00)$ vibrationally excited states of HCN near
  the Coriolis resonance involving the \J=8 to 12 rotational levels.
  Prominent laser lines detected in the laboratory are indicated by
  arrows along with their frequencies.  The bold arrow indicates the
  $(04^00)$, \J=9--8 transition discussed in the present paper.}
    \label{fig:fig2}
  \end{center}
\end{figure}

Fig.~2 shows HCN energy levels around the \J=10 and \J=9 rotational
states within the $(11^{1}0)$ and $(04^{0}0)$ vibrational ladders.
Fortuitously, several of the rotational levels in the $(11^{1}0)$
state are very close in energy to levels with identical \J\ in the
$(04^{0}0)$ state. This results in rotation-vibration interactions,
i.e. Coriolis coupling, and non-vanishing
transition probabilities for cross-ladder transitions (Lide \&\ Maki 1967).

To understand the occurence of laser action in these cross-ladder
and connected transitions, we note that
the (100) state is, practically 
speaking, metastable (Herzberg 1945),
with vibrational transitions from the (100)
to the (000) ground state being $\sim 2$ -- 3 orders of magnitudes
slower than
(010)--(000) and 
(001)--(000) transitions
(Kim \&\ King 1979).
In the laboratory, HCN molecules produced in a gas discharge
are distributed over the various vibrational states, most of which
[including the $(04^{0}0)$ state] will depopulate quickly by
spontaneous emission directly to the ground state or by cascading down the
vibrational ladder.  However, the metastable (100) state, 
and all combination states building upon it, such as
the $(11^{1}0)$ state, will become overpopulated relative to other
vibrational states, including the $(04^00)$ state.
A consequence of this is laser action in the
\J=11--10 and 10--9 cross-ladder transitions, which in turn 
causes the $(04^00)$, \J=10 and \J=9 states to be overpopulated and the
$(11^{1}0)$, \J=11 and 10 to be underpopulated, resulting
in laser action in the $(04^00)$, \J=10--9 and 9--8 as well as in the
$(11^{1}0)$, \J=12--11 and 11--10 transitions.

It is interesting to note that of the more than 120
cosmic maser and laser transitions known so far, 
the HCN $(04^00)$, \J=9--8 line is, apart from 
the $(J,K) = (3,3)$ inversion line of ammonia (NH$_3$, Mangum 
\&\ Wootten 1994), the only one that has been reported to show maser action in
an astronomical object as well as in the laboratory.
In the case of the NH$_3$ (3,3) maser, inversion in the laboratory 
(Gordon, Zeiger, \&\ Townes 1954) is 
achieved by completely different means than in interstellar space
(Flower, Offer, \&\ Schilke 1990). In contrast, quite remarkably, 
the HCN laser might be produced in IRC+10216's circumstellar envelope 
by essentially the same mechanism as in a gas discharge call. 
This will be discussed in 
\S3.3 after a we have described the environment in which the laser arises.

\subsection{Vibrationally excited HCN in IRC+10216's innermost envelope} 

Since the (0$4^0$0), \J=9--8 line is emitted from a very highly
excited vibrational state of the molecule, we 
try to characterize
the excitation conditions of hot HCN in IRC+10216's inner envelope by
using our own multi-line observations as well as data from the literature.

The first observations of vibrationally excited HCN in
IRC+10216 were made by Ziurys \& Turner (1986), who detected
the \J=3--2 transition in the $(01^{1_c}0)$, $(01^{1_d}0)$, and
$(02^{0}0)$ states and find
the levels to be thermally populated.  Guilloteau, Omont, \&\ Lucas (1987)
report strong maser emission in the $(02^{0}0)$, \J=1--0 transition 
toward the carbon star CIT 6. Weaker maser emission in this line is found
toward other carbon-rich stars (Lucas, Guilloteau, \&\ Omont 1988) including
IRC+10216 (Guilloteau et al. 1987). 
Interferometric observations toward the latter star
suggest that the maser action arises from within a region of radius
0\farcs 25, or $\approx 10~r_\star$ (Lucas \&\ Guilloteau 1992).
Strong maser action in the
$(01^{1_c}0)$, \J=2--1 transition was found toward IRC+10216 by
Lucas \& Cernicharo (1989), who
also observed thermally excited lines from the $(01^{1_d}0)$,
$(02^{0}0)$ and $(02^{2}0)$ states. 

Rotational lines from the (100) and (001) states of HCN were first detected by 
Groesbeck, Phillips, \&\ Blake (1994), who observed the \J=4--3 transition
in these as well as in several of the (01$^\ell$0) and (0$2^\ell$0) 
states.
Using the Infrared Space Observatory (ISO) Cernicharo et al. (1996) 
found emission in these  vibrational states from the
\J=18--17 up to the \J=48--47 rotational lines.
These observations indicate a vibrational temperature of 1000~K.        
     
\begin{figure}[htbp]
  \begin{center}
    \includegraphics[scale=0.5,angle=-90]{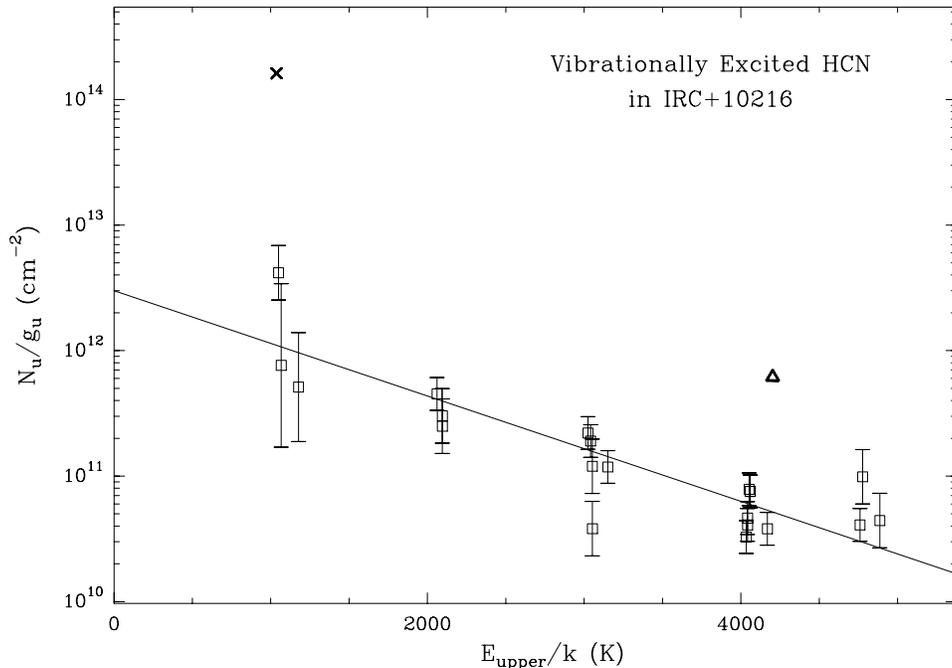}
\caption{Vibration--rotation Boltzmann plot for HCN,
  comprising our data and the (020) data from Groesbeck et al. 1994.
  The entries for the $(01^{1_c}0)$ maser and $(04^{0}0)$ laser lines
  are marked as an ``x'' and a triangle, respectively, and are not
  used in the fit.}
  \end{center}
\end{figure}
      
We complemented the data from the literature by using the CSO to
observe various other thermally excited lines from vibrationally
excited states with energies comparable to or even exceeding that of
the (0$4^0$0), \J=9--8 line. Details of these observations will be
given in a future publication.  Using the complete set of data we
construct the vibrational-rotational excitation diagram shown in
Fig.~3, for which we assume a point-like source and ``normalize'' the
intensities, which were measured with different resolutions, to a
11\arcsec\ FWHM beam. Clearly, the abscissa values of the data points
representing the $(01^{1_c}0)$, \J=2--1 maser and our $(04^00)$,
\J=9--8 laser are much higher than the thermal equilibrium values
expected for these lines.  Excluding those points, a least square fit
yields a vibrational temperature of 1000~K, similar to the value found
by Cernicharo et al. (1996). Assuming that the brightness temperature
of the thermally excited lines in the $(04^{0}0)$ state is equal to or
smaller than this number, we find an lower limit of 0\farcs 08, or
$\approx 3.5 r_\star$ for the radius of the emitting region.  This is
comparable to the upper limit found interferometrically for the radius
of the $(02^{0}0)$, \J=1--0 emission region and is
%
%
close to the  dust condensation radius, which according to the model
calculations of Groenewegen (1997) is at
4.5~$r_\star$ at a temperature of
1075~K. The high excitation vibrationally excited 
thermal lines have, in contrast to the laser line, symmetric profiles
covering a velocity range that is very similar to that of
the laser line, i.e.\ much narrower than that of lines from the
vibrational ground state. We therefore conclude that the
HCN laser line is formed at the dust condensation radius at a temperature 
of 1000~K.

Both the laser discussed here and the $(01^{1_c}0)$, \J=2--1 and
$(02^{0}0)$, \J=1--0 masers exhibit enhanced emission at
velocities that are blueshifted relative to the stellar velocity.
If a systematic outflow has already started in the part of the 
envelope giving rise to the HCN masers, one would expect
the blueshifted emission to arise from between the star and the observer,
allowing for the possibility that its greater intensity is due to
amplication of background emission from the stellar photosphere.
A significant
reduction of the redshifted emission due to geometrical blocking 
by the star can be excluded if the size of the HCN emission region 
has dimensions of the order discussed above. On the other hand, we know 
from Very Long Baseline Interferometry of SiO masers in O-rich Mira stars that 
red- and blue-shifted SiO maser emission does not necessarily arise 
from, respectively, the back and front parts of the
circumstellar shell; instead ring-like emission
distributions are observed, indicating tangential amplification
(Diamond et al. 1994).
Recent high resolution infrared observations in the $K'$ band 
(Weigelt et al. 1998) indicate that the dust shell of IRC+10216 is 
extremely clumpy and asymmetric to a radius of $\approx 0\farcs 2$ or
$9~r_\star$, 
which is comparable to the size of the dust formation and HCN laser/maser 
region. Maser pumping by such a highly anisotropic infrared field would
certainly  result in an inhomogenous emission distribution giving
rise to an asymmetric line profile. Very high resolution
interferometric observations with future instruments such as the Atacama Large
Millimeter Array (ALMA) are needed to image the maser emission and
study its dynamics.

\subsection{Pumping Schemes}

A direct radiative pump from the ground state into the $(11^10)$ state
does not seem likely, since in that case the other 
vibrational states would be preferentially 
populated by virtue of their higher transition 
probabilities. We were unable to check for possible line overlaps between
infrared transitions from HCN and other molecules abundant in IRC+10216,
such as CO, and pumping by that mechanism remains viable in principle.
Collisional pumping might also be possible, since the critical density
(i.e., the ratio of the Einstein A-coefficient
to the collisional excitation coefficient) of the $(100)$ stack, which 
includes, among others, the $(11^10)$ combination state, 
is lower than that of the other vibrational states,
if we assume comparable collision rates. A more
detailed analysis of this mechanism requires knowledge of 
collisional excitation coefficients.

Most interestingly, chemical pumping, which has been invoked to explain 
the laboratory HCN lasers (see, e.g., Chantry 1971; Pichamuthu 1983), 
might also be at work in IRC+10216.  
The basic idea is that the HCN molecules are formed in random
vibrational states, and a population inversion is
achieved in the $(11^10)$/$(04^00)$ system because of the wide difference
in the radiative decay rates of the different vibrational levels.
For this mechanism to work, the number of HCN creation events per time unit
should exceed the number of HCN laser photons emitted by a considerable
amount.
HCN is formed in thermochemical equilibrium (see Tsuji 1964), where
the reaction of CN with \HTW\ is likely to be a major channel. 
In the following we assume that HCN formation proceeds at a typical 
rate of $10^{-10}$ s$^{-1}$ in a
shell between radii of $2\times 10^{14}$
and $3\times 10^{14}$~cm, i.e. just
interior to the dust formation radius.  If we further assume a CN abundance
of $10^{-11}$ (the value at 1000~K; Lafont, Lucas \& Omont 1982)
and an \HTW\ density of $10^{12}$ cm$^{-3}$, we calculate $8\times
10^{45}$ HCN creation events per second. 
From our spectrum we determine an integrated line flux of 
6100~Jy~\kms, which corresponds to an isotropic photon luminosity
of $7\times10^{43}$ s$^{-1}$. Given that any alternative production mechanism 
would only increase the number of creation events, it appears that
chemical pumping seems possible, although the uncertainties in the
numbers used are considerable.

Finally, we note that if the circumstellar 805 GHz laser results from 
overpopulation of the $(11^{1}0)$ state 
and $(11^10)$--$(04^00)$ cross-ladder transitions, as is highly likely,
one would also expect the other transitions 
of this Coriolis-coupled system, like in the laboratory, to show
strong laser action toward IRC+10216. Of these (see Fig. 2), the 
$(11^10)$--$(04^00)$, \J=10--9 cross ladder transition at 890.76 GHz
and the 894.31 GHz \J=10--9 line within the  $(04^00)$ state are, in principle,
observable from high, dry mountain sites, 
but outside the frequency range of the 
receiving system currently 
available to us. The $(11^10)$--$(04^00)$, \J=11--10 and
$(11^10)$, \J=11--10 transitions at 964.31 and 967.96 GHz, respectively, 
should be observable with the Stratospheric Observatory for Infrared 
Astronomy (SOFIA).
\vskip 0.2truein 

We would like to thank Tom Phillips, the director of the CSO, for 
granting us observing time, Darrel Dowell for help with some of the
observations, and Mark Reid for comments on the manuscript.
Work at the Caltech Submillimeter Observatory
is supported by NSF contract AST 96-15025.


\begin{references}

\reference{}Avery, L. W. et al. 1992, ApJS, 83, 363

\reference{}Cernicharo, J., Bujarrabal, V., \&\ Santar{\'e}n, J. L. 1993, ApJ,
407, L33

\reference{}Cernicharo, J., et al. 1996, A\&A, 315, L201

\reference{}Cernicharo, J., Gu{\'e}lin, M., \&\ Kahane, C. 1999, A\&A, in press

\reference{}Chantry, G. W. 1971, Submillimetre
    Spectroscopy, (London: Academic Press), p. 241

\reference{}Crosas, M., \&\ Menten, K. M. 1997, ApJ, 483, 913

\reference{}Danchi, W. C., Bester, M., Degiacomi, C. G., Greenhill, L. J., 
\&\ Townes, C. H. 1994, AJ, 107, 1469


\reference{}Diamond, P. J., Kemball, A. J., Junor, W., Zensus, A., Benson, J.,
\&\ Dhawan, V. 1994, ApJ, 430, L61

\reference{}Flower, D. R., Offer, A., \&\ Schilke, P. 1990, MNRAS, 244, 4p

\reference{}Gebbie, H. A., Stone, N. W. B., \&\ Findlay, F. D. 1964, Nature,
202, 685

\reference{}Gordon, J. P., Zeiger, H. J., \&\ Townes, C. H. 1954, Phys. Rev.,
95, 282 

\reference{}Greenhill, L. J., Colomer, F., Moran, J. M., Backer, D. C., 
Danchi, W. C., \&\ Bester, M. 1995, ApJ, 449, 365

\reference{}Groenewegen, M. A. T. 1997, A\&A, 317, 520

\reference{}Groenewegen, M. A. T., van der Veen, W. E. C. J., \&\ 
Matthews, H. E. 1998, ApJ, 338, 491

\reference{}Groesbeck, T. D., Phillips, T. G., \&\ Blake, G. A. 1994,
 ApJS, 94, 147

\reference{}Guilloteau, S., Omont, A., \&\ Lucas, R. 1987, A\&A, 176, L24

\reference{}Habing, H. J. 1996, A\&AR, 7, 97

\reference{}Herzberg, G. 1945, Molecular Spectra and 
  Molecular Structure, II. Infrared and Raman Spectra of Polyatomic
  Molecules (Princeton: Van Nostrand), p. 279

\reference{}Hocker, L. O., \&\ Javan, A. 1967, Phys. Lett., 25A, 489

\reference{}Kawaguchi, K., Kasai, Y., Ishikawa, S., \&\ Kaifu, N. 1995, 
PASJ, 47, 853

\reference{}Kim, K., \&\ King, W. T. 1979, J. Chem. Phys., 71, 1967

\reference{}Kooi, J. W., Pety, J., Bumble, B., Walker, C. K., LeDuc, H. G.,
Schaffer, P. L., \&\ Phillips, T. G. 1998,
IEEE Trans. Microwave Theory and Technique, 46, 151

\reference{}Kneub{\"u}hl, F. K., \&\ Sturzenegger, C. 1980, in 
Infrared and Millimeter Waves Vol. 3, ed.
K. J. Button (New York: Academic Press), 219

\reference{}Lafont, S., Lucas, R., \&\ Omont, A. 1982, A\&A, 106, 201

\reference{}Lide, D. R. Jr., \&\ Maki, A. G. 1967,  Appl. Phys. Lett.,
  11, 62

\reference{}Lucas, R., \&\ Cernicharo, J. 1989, A\&A, 218, L20

\reference{}Lucas, R., \&\ Gu{\'e}lin, M. 1999, to appear in
Asymptotic Giant Branch Stars, ed. A. Lebre, T. Le Bertre, \&\ C. Waelkens

\reference{}Lucas, R., \&\ Guilloteau, S. 1992, A\&A, 259, L23

\reference{}Lucas, R., \&\ Guilloteau, S., \&\ Omont, A. 1988,
  A\&A, 194, 230

\reference{}Mangum, J. G., \&\ Wootten, A. 1994, ApJ, 428, L33

\reference{}Mathias, L. E. S., Crocker, A., \&\ Wills, M. S. 1965,
Electronics Lett., 1, 45

\reference{}Menten, K. M., \&\ Young, K. 1995, ApJ, 450, L67

\reference{}Pichamuthu, J. P. 1983, in 
Infrared and Millimeter Waves Vol. 3, ed.
K. J. Button (New York: Academic Press), 165

\reference{}Pickett, H. M., Poynter, R. L., Cohen, E. A., Delitsky, M. L.,
Pearson, J. C., \&\ M{\"u}ller, H. S. P. 1998, J. Quant. Spectrosc. \&\ Rad. 
Transfer, 60, 883 (see also http://spec.jpl.nasa.gov)

\reference{}Strelnitski, V., Haas, M. R., Smith, H. A., Erickson, E. F., 
Colgan, S. W. J., \&\ Hollenbach, D. J. 1996, Science, 272, 1459

\reference{}Tsuji, T. 1964, Ann. Tokyo Astron. Obs., 9, 1

\reference{}Turner, B. E. 1987a, A\&A, 182, L15

\reference{}Turner, B. E. 1987b, A\&A, 183, L23

\reference{}Weigelt, G., Balega, Y., Bl{\"o}cker, T., Fleischer,
  A. J., Osterbart, R., \&\ Winters, J. M 1998, A\&A, 333, L51              


\reference{}Winters, J. M., Dominik, C., \&\ Sedlmayr, E. 1994, 
A\&A, 288, 255

\reference{}Ziurys, L. M., \&\ Turner, B. E. 1986, ApJ, 300, L19

\end{references}
\end{document}